\documentstyle[12pt]{article}

\begin{document}

\title{{\LARGE {\bf Mechanism of the Atmospheric Ball Lightning Using the Triple
Beltrami Equation}}}
\author{{Ana M\'arcia Alves Taveira and Paulo Hiroshi Sakanaka} \\
{\small Instituto de F\'{\i}sica ``Gleb Wataghin'', UNICAMP, C.P. 6165}\\
{\small 13083-970, Campinas, SP, Brazil}}
\maketitle

\baselineskip=20pt \oddsidemargin =0.cm


{\Large {\bf {Abstract}}}

{\ Ball lightning, also known as fire ball, is a luminous globe which occurs in the course of  a
thunderstorm. It has been the object of investigation by figures  in
science since the early nineteenth century. The difficult and long-standing
problem of ball lightning has attracted few meteorologists or atmospheric
scientists. Rather, physicists constitute the larger part of the company
studying the atmospheric fire balls. }

Taking as model, the two fluid plasma consisting of electrons and one 
species ions, for the fire balls physics and considering that  the plasma
flow is a finite quantity, we can derive the equation of  relaxed energy
state, maintaining the helicity constant, in the  form of triple Beltrami
equations for the magnetic field:

\[
s \nabla\times\nabla\times\nabla\times \vec{B}  + p \nabla\times\nabla\times 
\vec{B} +  q \nabla\times \vec{B} + r \vec {B} =0 
\]
where $\vec{B}$ is the magnetic field, $s$, $p$, $q$, and $r$ are constants
to be determined through boundary conditions. This equation is coupled with an 
equation which describes the hydrodynamic vortex.

When $s=p=0$, we have the Taylor relaxed state of plasma [1], without the
fluid  flow.  In the case of electron-ion plasma, both $s$ and $p$ are
non-zero quantities. In particular, when electron mass is neglected the
results is $s=0$ and it describes the Double Beltrami relaxed minimum energy 
state as derived by S. M. Mahajan and Z. Yoshida [2]. Its solution is a
spheromak type solution.  The problem of the formation of an isolated
luminous mass in the sky and the moderate persistence of the resulting form
combined with observations of ball lightning describing hollow globes,
surface coronas and rapid rotation led to theories depicting ball lightning
as a vortex. We explore the solutions of this  equation using the method of
Chandrasekhar-Kendall [3] eigenfunctions  and appropriate boundary
conditions similar, but more extensive, to the description given by  Shukla, Dasgupta and
Sakanaka [4]. We will show solutions which might  explain the fire ball
configurations. \vskip0.2in


\leftline {\bf\large 1. Introduction} \vskip0.2in

Ball lightning is a luminous globe which occurs in the course of  a
thunderstorm.
It has been the object of investigation by well known scientists 
since the early nineteenth century. Although its appearance is very haphazard
and far less frequent than ordinary linear lightning, it is a source of
astonishment when the fire ball is observed in the lower atmosphere, often
entering dwellings while floating at a leisurely pace. By its random
appearances, it has eluded the measurements with the scientific instruments. 
Even photographic captures of fire balls are rare.
However, from the qualitative reports of eye-witnesses reports (now it numbers in the
thousands, particularly from collections from the Soviet Union) the general
properties can be deduced despite the wide variability found in the reports.

Ball lightning makes appearance as flame globe, typically in orange, red-orange
or intense white, and less often in blue, green or yellow. Its dimensions is usually 25-30
cm in diameter, but much smaller or much larger dimensions are also notified in
the reports (from 1 cm to 10 m and larger). The color of the ball lightning
would depend on the current  in the channel as in the experiments in which a
weak current  gives a bluish glow while increasingly stronger currents gives
dark red, brick red, orange red, and finally white. Occasionally, the balls
move against the direction of prevailing winds and penetrate window panes
without making a hole in the glass; it is these properties that Ohtsuki and
Ofuruton report from their experimental discharges.

The difficult and long-standing problem of ball lightning has attracted only
few meteorologists or atmospheric scientists. Rather, physicists constitute
the larger part of the company studying the atmospheric fire balls. Numerous
theories have been put forward, covering chemical, electrical, nuclear and
relativistic models. New states of matter have been proposed to explain the
unusual properties reported for ball lightning.

The problem of the formation of an isolated luminous mass in the sky and the
moderate persistence of the resulting form combined with observations of
ball lightning describing hollow globes, surface coronas, and rapid rotation
led to theories depicting ball lightning as a vortex.

The generation of small vortexes constituting ball lightning  in the
atmosphere by whirl winds, cyclones, or tornadoes was  suggested following
numerous observations of  fire balls in connection with a tornado which
appeared at  night in France in 1890.

The possibility of a close relationship between the formation  of ball
lightning and tornadoes has been suggested,  The combined action of
electrical and hydrodynamic forces in  generating ball lightning as a
highly ionized vortex was  specifically adduced in general terms in 1905.

The magnetic field generated by rotation of charged particles in the  ring,
evidently in helical paths in its cross section as well as in  circular
paths around the ring, was suggested to assist in confinement. The discharge
of a large current through a fine wire ring  was presented as a method of
generating such a vortex.

The vortex theories provide a very direct explanation of  the numerous
observations indicating rotation of ball lightning.  The formation of
luminous globes of this type can be readily  ascribed to either the role of
a preliminary linear lightning  flash or the hydrodynamic action of a
whirl wind.

Some vortex theories may be classified equally well as electrical  discharge
theories and, especially in the most recent examples,  as plasma theories.
None of the studies of ball lightning as a  vortex, however, has
conclusively dispelled by detailed consideration  the serious difficulties
on which other theories have founded,  such as the continued luminosity of
the balls for long periods  while they travel inside structures. If the
vortex is to be  an isolated, self-contained sphere, its energy would be 
expended in viscous drag and turbulence in a very short time.

We explore the possibilities of relaxed states of plasma for atmospheric
 plasma with positive charged particles, and electrons in the form of
Triple Beltrami equations. It is shown that the associated velocity field
has the similar vortex structure as in the magnetic field. Solutions of this
equation is obtained using the method of Chandrasekhar-Kendall
eigenfunctions and appropriate boundary conditions similar, but more extensive, to the
description given by Shukla, Dasgupta and Sakanaka [2] . We will show
solutions which might explain the ball lightning configurations.


\vskip0.2in \leftline {\large \bf Theory} \vskip0.2in

We consider a warm, homogeneous, electron-ion plasma. The dynamics of low
phase velocity e-i plasma are governed by the electron and ion momentum
equations, which are, respectively,

$$
\frac{\partial {\bf {v_{e}}}}{\partial t}+\frac{1}{2}\nabla {\bf {v_{e}}}%
^{2}-{\bf {v_{e}}}\times \nabla \times {\bf {v_{e}}}=-\frac{e}{m_{e}}\left[ 
{\bf {E}}+\frac{1}{c}{\bf {v_{e}}}\times {\bf {B}}\right] -\frac{1}{%
m_{e}n_{e}}\nabla p_{e}\eqno(1)
$$

\noindent and

$$
\frac{\partial{\bf {v_i}}}{\partial t}+\frac{1}{2} \nabla{\bf {v_i}}^2-{\bf {%
v_i}}\times\nabla\times{\bf {v_i}} = \frac{e}{m_i}\left[{\bf {E}}+ \frac{1}{c%
}{\bf {v_i}} \times{\bf {B}}\right]-\frac {1}{m_i n_i}\nabla p_i \eqno(2)
$$

\noindent supplemented by Faraday's law

$$
\frac{\partial {\bf {B}}}{\partial t}=-c\nabla \times {\bf {E}},~~\eqno(3)
$$

\noindent and~Amp{e}re's~law

$$
\nabla\times{\bf {B}} = \frac{4\pi}{c}{\bf {J}} = \frac{4\pi}{c}ne({\bf {v_i}%
} -{\bf {v_e}})\equiv \frac{8\pi}{c}ne{\bf {U}} \eqno(4)
$$

\noindent where ${\bf {v_e}} $ $({\bf {v_i}})$ is the electron (ion) fluid
velocity, ${\bf {U}} =\frac{1}{2}({\bf {v_i}} -{\bf {v_e}})$, and, ${\bf {E}}$ $
({\bf {B}})$ is the electric (magnetic) field, $n_e$ ($n_i$) is the uniform electron
(ion) number density (given by the continuity equation), $e$ is the
magnitude of the electron charge, $m_{e}$ ($m_i$) is the electron (ion) mass, and 
$p_e$  $(p_i)$ is the scalar electron (ion) pressure. The fluid velocity ${\bf {%
V}}$ for e-i plasma is defined by,

$$
{\bf {V}} = \frac{{m_i}{\bf v_i}+{m_e}{\bf v_e}}{{m_i}+{m_e}} \simeq {\bf v_i}+{\mu ~\bf v_e}%
, ~~\mu=\frac{m_e}{m_i} \eqno(5)
$$

\noindent one gets

$$
{\bf {v_i}} = {\bf {V}} + 2 \mu {\bf {U}},~~~~ {\bf {v_e}} = {\bf {V}} - 2 {\bf 
{U}} \eqno(6)
$$

\noindent From equation (4), we can write,

\[
{\bf {U}} = \frac{c}{8\pi ne}\nabla\times{\bf {B}}
\]

\noindent Using this, the electron and ion velocities can be expressed in
the normalized form

$$
{\bf {v_i}}={\bf {V}}+\sqrt{\mu}~\nabla\times {\bf {B}},~~{\bf {v_e}} = {\bf {V%
}}  - \frac{1}{\sqrt{\mu}}\nabla\times{\bf {B}}\eqno(7)
$$

\noindent We normalize the variables and to express the momentum equations,
we start with (1) and (2) and after taking ``curl'' of both sides, in a
dimensionless form; the magnetic field ${\bf {B}}$ is normalized to some
arbitrary $B_0$, the velocity to $V_0 = B_0/\sqrt{4\pi {n_e}{m_i}}$, length
and time by skin depth for e-i plasma $\lambda = c/\omega_{p_e} = c/\sqrt{%
4\pi {n_e}e^2/{m_e}}$ and inverse of cyclotron frequency, $\tau_c = ({m_e}%
c/eB_0)$ respectively.  Taking ``curl'' of both sides of equations (1) and
(2), and using equation (7) and the Faraday's law, equation (3), we get the
momentum equations for electron and ion in normalized variables. Introducing
a pair of generalized vortices, ${\bf {\Omega_i}}, ~{\bf {\Omega_e}}$

$$
{\bf {\Omega_i}} = \nabla\times{\bf {v_i}}+ \sqrt{\mu}~{\bf {B}};~~ {\bf {%
\Omega_e}} = \nabla\times{\bf {v_e}}- \frac{1}{\sqrt{\mu}}~{\bf {B}} \eqno(9)
$$

\noindent and effective velocities, ${\bf {U_i}}, ~{\bf {U_e}}$ (these
velocities are the normalized electron and ion velocities), where

$$
{\bf {U_e}} = {\bf {V}} -\frac{1}{\sqrt{\mu}} ~\nabla\times{\bf {B}}, ~~ {\bf 
{U_i}} = {\bf {V}} + \sqrt{\mu}~\nabla\times{\bf {B}}\eqno(10)
$$

\noindent the electron and ion momentum equations can be put in a symmetric
form,

$$
\frac{\partial{\bf {\Omega_j}}}{\partial t} - \nabla\times({\bf {U_j}}\times%
{\bf {\Omega_j}}) = 0,~~~~ (j=i, e)\eqno(11)
$$

\noindent The above equations show the effects of the coupling of magnetic
field and flow in an exact form. We can look for a simple equilibrium
solution of the above equation, the simplest equilibrium is obtained as $%
{\bf {U_j}}\parallel{\bf {\Omega_j}}, ~~(j=1,2)$. Thus we get,

$$
{\bf {U_e}} = {\bf {V}} - \frac{1}{\sqrt{\mu}} ~\nabla\times{\bf {B}} = a_1
(\nabla\times ({\bf {V} - \frac{1}{\sqrt{\mu}}~ \nabla\times{B}) - \frac{1}{%
\sqrt{\mu}~{B}})}\eqno(12)
$$

\noindent and,

$$
{\bf {U_i}} = {\bf {V}} + \sqrt{\mu}~\nabla\times{\bf {B}} = a_2
(\nabla\times ({\bf {V} + \sqrt{\mu}~\nabla\times{B}) + \sqrt{\mu}~{B})}\eqno(13)
$$

\noindent where, $a_1$ and $a_2$ are two arbitrary constants, to be
determined by the physics of the problem (such as boundary conditions, etc).
The above two equations can be combined, after eliminating either ${\bf {V}}$
or ${\bf {B}}$.

The equation for $\bf B$ can be written,

{
$$
\nabla\times\nabla\times\nabla\times{\bf {B}} + p\nabla\times\nabla\times%
{\bf {B}} + q\nabla\times{\bf {B}} + r{\bf {B}} = 0\eqno(14)
$$
}

{ \noindent }where{ \ $p,~q,~r$ }are constants, given in terms of%
{ \ $a_{1}$ }and{ \ $a_{2}$: $p=-\frac{a_{1}+a_{2}}{a_{1}a_{2}}$%
, ~~ $q=\frac{1+{a_{1}}{a_{2}}}{{a_{1}}{a_{2}}}$}, and{ \ $r=\frac{-1}{%
a_{2}}$ }

It may be pointed out that the solution spectrum of equation (14) is much
wider and richer that those obtained from the solution of the corresponding
equation for electron ion fluid. The general solution of eqn.(14) can be
obtained as a linear superposition of the
Chandrasekhar-Kendall eigenfunctions, which are the eigenfunctions of the
``curl'' operator, i.e.
{
\[
\nabla\times{\bf {B}} = \lambda{\bf {B}}
\]
}
\indent This simple solution shows some remarkable properties of e-i plasma. The
pressure profile, peaking at the axis, shows the self-confining properties
of this equilibrium state. Also, the axial magnetic field reverses at the
edge, showing the field-reversal feature, like that seen in the Reversed
Field Pinch (RFP). Thus, this work demonstrates that e-i plasma, under
appropriate circumstance can behave like a RFP, but with a confining
pressure. We believe that this could open a new direction in activities of
e-i plasma, like investigation of vortex like structures (i.e., spheromak
type of closed field lines) and many other new and interesting physics. 

{\large 
\vskip 0.2in \leftline {\bf Acknowledgments} \vskip 0.2in }

This work was partially supported by the Brazilian agencies FAPESP, CNPq and
FINEP-PADCT.{\large \ }

{\large \vskip0.2in }

{\large {\small 
}}


\begin{thebibliography}{9}


\bibitem{taylor74}  {\large {\small J.~B. Taylor. \newblock {\em Phys. Rev.
Lett.}, 33:139, (1974). }}

\bibitem{mahajan98}  {\large {\small S.M.~Mahajan and Z.~Yoshida, Phys. Rev.
Lett., {\bf 81}, 4863 (1998). }}

\bibitem{chandra57}  {\large {\small S.~Chandrasekhar and P.C.~Kendall,
Astrophys. J., {\bf 126}, 457 (1957). }}

\bibitem{saka2000}  {\large {\small \ P.K.~Shukla, B.~Dasgupta, and
P.H.~Sakanaka,Phys. Lett.,{\bf A269}, 144 (2000). }}
\end{thebibliography}
\end{document}